\documentstyle[12pt]{article}
\begin{document}
\baselineskip=24pt
\begin{center}
{\large Spin liquid in 3D Kondo lattice. \\
High temperature regime}
\\
K.A.Kikoin, M.N.Kiselev, A.S.Mishchenko
\\
{\it RRC Kurchatov Institute, Moscow 123182, Russia}
\end{center}
\begin{abstract}
The mechanism explaining the key role of 
AFM correlations in formation of the heavy fermion state is offered. 
It is shown that in the case of   
$(T^*-T_N)/T_N \ll 1$ the critical spin fluctuations transform  
the mean-field second-order transition to RVB state 
into the crossover from high-temperature paramagnetic behavior of 
localized spins to strongly correlated spin liquid with quasi itinerant 
character of susceptibility. Thus the spin liquid state {\it by its origin} 
is close to magnetic instability, so either short-range or long-range AFM 
order should arise at low T.
\end{abstract} 
\mbox{} \\
K.A.Kikoin \\
RRC Kurchatov Institute, Moscow 123182, Russia\\
tel: (7)(095)1967584; \\ fax: (7)(095)1965973;
\\ e-mail: kikoin@kurm.polyn.kiae.su

\newpage
1. It was shown in \cite{1} that the spin-liquid state of resonating valence 
bond (RVB) type can be stabilized by the Kondo stattering against the 
antiferromagnetic (AFM) ordering in 3D Kondo lattices at high enough 
temperatures $T^*>T_K$, provided the Kondo temperature $T_K$ is close to 
the mean-field (MF) Neel temperature $T_N$. This stabilization is due to the 
fact that the Kondo scattering screens dynamically the local moments and, 
thus, suppresses the magnetic order,  whereas 
the spin-liquid correlations which result in the singlet RVB are 
left intact by this scattering. The theory was based on the MF 
description of RVB state which, as is known \cite{2}, results in second-order 
type transition from paramagnetic state to RVB phase. However, since the 
transition to the spin liquid state takes place close to the point of AFM 
instability, the critical fluctuations influence essentially the character 
of spin liquid transition. The description of this influence is the main 
subject of the present paper. 
\mbox{}\\
2. We start with a standard Kondo lattice Hamiltonian 
\begin{equation}
H_{eff}=\sum_{\bf{k} \sigma}\varepsilon_k 
c^+_{\bf{k} \sigma}c_{\bf{k}\sigma} +
J_{sf}\sum_{\bf{i}}c^+_{{\bf i}\sigma}c_{\bf{i}\sigma'}
f^+_{{\bf i}\sigma'}f_{{\bf i}\sigma} 
\label{eq.1}
\end{equation}
Here $\varepsilon_k$ is the band level of conduction electron
$c_{\bf{k}\sigma}$ , and the operator of localized f-spin represented 
via fermionic operators, 
$S_{\bf i} = 
\frac{1}{2}f^+_{\bf{i} \sigma} \hat{\sigma} f_{\bf{i} \sigma'}$,  
where $\hat{\sigma}$ is the Pauli matrix.
At $T>T_K$ the non-crossing approximation for the 
Kondo scattering processes is valid, and this Hamiltonian can be transformed 
into effective RKKY-type Hamiltonian for spin variables only, 
\begin{equation}
H_{RKKY}=I_{ss}\sum_{\bf ij}f^+_{{\bf i}\sigma}f_{\bf{j}\sigma}
f^+_{{\bf j}\sigma'}f_{{\bf i}\sigma'} 
\label{eq.2}
\end{equation}
where $I_{ss}\sim \tilde{J}_{sf}^2(T^*)/\varepsilon_F$ 
is the indirect RKKY exchange interaction with the sf-vertices 
$\tilde{J}_{sf}(T^*)$ enhanced by the Kondo scattering which is 
taken into account
in a high-temperature approximation of perturbation theory 
\cite{1}. The Kondo processes are "quenched" at some temperature $T^*>T_K$ 
which characterizes the onset of spin-liquid RVB state described by the 
variables $b_{ij}=\sum_{\sigma}f^+_{{\bf i}\sigma}f_{{\bf j}\sigma}$
under the constraint $\sum_{\sigma}f^+_{{\bf i}\sigma}f_{\bf{i}\sigma}=1$. 
If one introduces the MF parameter $\Delta =\langle b_{ij} \rangle$ in 
the Hamiltonian (\ref{eq.2}), the temperature $T^*$ is determined as 
$T^*=I_{ss}(2zN)^{-1}\sum_{\bf k}\varphi^2(\bf k)$, 
where $\varphi(\bf k)$ is a lattice structure factor with the 
coordination number $z$.
Just this temperature was shown in \cite{1} to become  higher then $T_N$ 
in a critical region $T_K\sim T_N$ of the Doniach state diagram. 
\mbox{}\\
3. Since the inequality $(T^*-T_N)/T_N \ll 1$ is valid for the MF 
solution, the closeness to AFM instability should be taken 
into account. This closeness enriches the phase diagram of 3D spin liquid 
in comparison  with the MF scenario of homogeneous RVB state formation 
\cite{1}. We consider here the Kondo lattice with AFM-type RKKY interaction 
for the nearest neighbors which could result in commensurate ordering 
with AFM wave vector ${\bf Q}$ at 
$T=T_N$ provided the RVB state was not realized at higher temperature $T^*$. 
This means that the denominator of the static susceptibility 
\begin{equation}
\chi_{\bf Q} = \chi_0(T)\left[1 - \chi_0(T)I({\bf Q})\right]^{-1}
\label{eq.4}
\end{equation}
is close to zero at $T\approx T^*$. Here  
$\chi_0(T)=\langle{\bf S}_{\bf Q}\cdot{\bf S}_{\bf -Q}\rangle_{\omega=0}=C/T$ 
is the Curie susceptibility of  free localized spin. 
Below $T^*$ the latter acquires dispersion, begins to deviate gradually 
from Curie law, and finally takes the form 
\begin{equation}
\chi_{0{\bf Q}}(T) = N^{-1}\sum_{\bf k}
\left(n_{\bf k} -n_{\bf k+Q})\right)\left(t_{\bf k+Q}-t_{\bf k}\right)^{-1}
\label{eq.5}
\end{equation}
where $n_{\bf k}$ is the Fermi distribution function for the RVB excitations 
which are characterized by the dispersion law $t_{\bf k}$. This deviation 
is shown schematically by the solid curves in Fig. 1. 
The zero-temperature limit of these curves can be estimated as 
$\chi_{0{\bf Q}}^{-1}(0)=\alpha T^* $ where $\alpha$ is the 
numerical coefficient which value depends on the character of phase 
transition. This value can be either lower or higher then $T_N=CI({\bf Q})$. 

In the first case (curve 1 in fig. 1) 
the point $\tilde{T}_N$ where $\chi_{0{\bf Q}}^{-1}$ crosses the dotted line, 
corresponds to AFM transition. However, 
the character of this transition differs 
from that of the localized spins. According to our scenario, the spin
subsystem in the Kondo lattice bypasses the magnetic instability at $T=T_N$ 
to order at  essentially lesser temperature $\tilde{T}_N$. However, within 
the interval $T_N>T>\tilde{T}_N$ the localized spins are transformed into 
spin liquid, and, as a result, the magnetic order reminds rather the itinerant 
AFM of conduction electrons with modulated spin density and 
small moments.      

Since the spinon band is always half-filled because of the constraint, the 
nesting condition with the same wave vector ${\bf Q}$
can be realized for some geometries of the Fermi surface. 
In this case $\chi_{0{\bf Q}}^{-1}$ logarithmically 
diverges, $\chi_{0{\bf Q}}\sim  \Delta^{-1}\ln{\Delta/T}$, so the 
SDW-type magnetism appears at $T_{SDW}$. 

Next possibility occurs when the curve $\chi_{0{\bf Q}}^{-1}$ does not 
intersect $T_N$ but comes close enough to it (curve 3). 
In this case we meet the 
peculiar situation when the paramagnon-type excitations can develop in the 
absence of itinerant electrons because the spin excitations has their own 
Fermi-type continuum in a spin liquid state. Then, the spin-fluctuation 
order which reminds that for itinerant electrons \cite{3} 
can occur at some temperature $T_{sf}$. 
If these fluctuations are 
too week to provide the long-range order, the short-range magnetic 
correlations characterized by the vector ${\bf Q}$ 
persist at low temperatures. 
Thus, we see that the model grasps the whole variety 
of peculiar magnetic states with tiny itinerant-like moments which are known 
for the heavy-fermion materials.
\mbox{}\\
4. The AFM spin fluctuations influence also the character of transition 
from the paramagnetic state to the RVB state. 
It is known that the second-order transition at $T=T^*$ is 
the artifact of the MF approximation which violates the 
gauge invariance of the Hamiltonian (\ref{eq.1}), and the real situation is 
that of crossover type. Our approach demonstrates that the real phase 
transition in Kondo lattice is {\it always} magnetic phase transition, and the 
spin-liquid-type correlations only change the character of this transformation.

To show this, we consider the MF self-energy 
$\Sigma_{\bf ij}$ of the temperature spinon Green's function 
$G_{\bf ij}(\tau)=
-\langle T_{\tau}f_{{\bf i}\sigma}(\tau)f^+_{{\bf j}\sigma}(0)\rangle$ 
corrected by the critical AFM fluctuations (fig. 2a). 
Here the lines stand for the Fourier-transformed $G_{\bf ij}$, 
full dots correspond to 
$I({\bf Q})$, the loop means 
spin susceptibility, $\chi_{\bf Q}(\varepsilon_n)$, whose static part  
is given by (\ref{eq.4}). The latter correction, 
$\delta\Sigma_{\bf p}$, inserts
retardation into $G_{\bf p}(\omega_m)$ and "opens" the system of equations 
which determine the spinon correlators for the influence of critical 
fluctuations. The correction $\delta\Sigma_{\bf p}$ 
corresponds to the mode-mode coupling term in the free energy functional, 
$\delta {\cal F}_{\bf Q}=J^2({\bf Q})M_{\bf Q}^{AFM}M_{\bf Q}^{RVB}$ 
where $M_{\bf Q}^X$ means the spin density fluctulations of localized 
$(X=AFM)$ and itinerant $(X=RVB)$ type. 

To find $\chi_{\bf q}(\varepsilon_n)$ for ${\bf q}$ close to 
${\bf Q}$  one should take into account that the RVB continuum exists
at temperature $T>T^*>T_N$ but still in the critical region of AFM 
instability  as a "virtual" continuum  of spinon 
particle-hole pairs with the gap $\Omega_0$ and nonzero damping $\gamma$. 
These excitations give both static ($\delta\chi_{0\bf q}$)  and  dynamical 
($\delta\chi_{\bf q}(\varepsilon_n)$) contributions 
in a simple spinon loop and, 
correspondingly, in the RPA equation for $\chi_{\bf Q}(\varepsilon_n)$. 
The vertex corrections in $\gamma/\Omega_0$ should be also taken into account 
(fig. 2b).
The static contribution is responsible for initial deviation of 
$\chi_{0\bf Q}(T)$ (fig. 1) from the Curie law, and 
the calculation of dynamics reminds formally that offered in \cite{4} for the 
2D Heisenberg model at finite temperatures. Our spinon-pair excitations with 
the gap play the same role as the Schwinger bosons in \cite{4}. So, 
we come to the similar result: the relaxation mode appears in 
$\chi_{\bf q}(\Omega)$ $(\Omega$ is the frequency analytically continued to 
the real axis),
\begin {equation}
\chi^{-1}({\bf q},\Omega) \sim 
\gamma + (qD\gamma^{-1/2})^2 - i\Omega
\label{6}
\end{equation}  
where $D$ is a sort of diffusion coefficient. 

In conclusion, we have found that the RVB-type excitations appear in the 
Kondo lattice at high temperature, first, as a relaxation mode in 
susceptibility due to closeness of $T^*$ to $T_N$. 
These excitations evolve into fermi-type particle-hole pairs with lowering 
temperature, and, eventually, the magnetic order of itinerant-like character 
arises at $\tilde{T}_N\ll T^*, T_K$. The relaxation regime is 
extended in $T$ in comparison with standard critical regime for the localized 
moments, and even can persist at $T\rightarrow 0$. 
The theory predicts the presence of inelastic neutron scattering 
at $\Omega \approx \Omega_0$ and ${\bf q} \approx {\bf Q}$.   
The small magnetic moments correspond to modulation of spin-liquid 
density. This picture correlates with experimental observations for 
CeRu$_2$Si$_2$ and CeCu$_6$ \cite{5}. 

The support of the grants NWO 5-16501, INTAS 93-2834 and RFBR-95-02-04250 
is acknowledged. 

\vspace{5cm}
\newpage
{\bf Figure captions}
\vspace{1cm}\\
Figure 1. Static magnetic susceptibility of RVB spin liquid 
\\
Figure 2. Self energy part of spinon Green's function (a)
and dynamic susceptibility with vertex corrections (b)
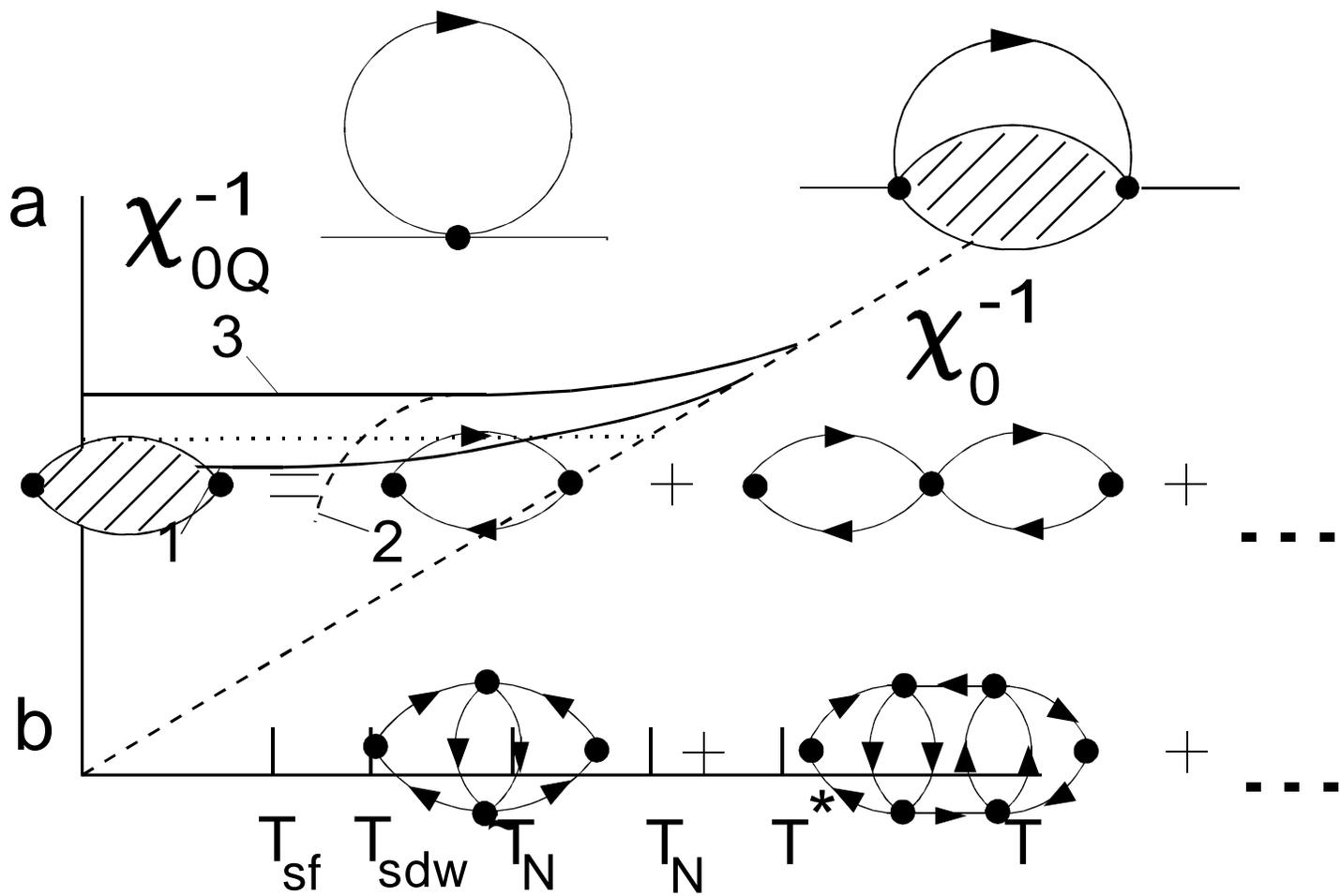
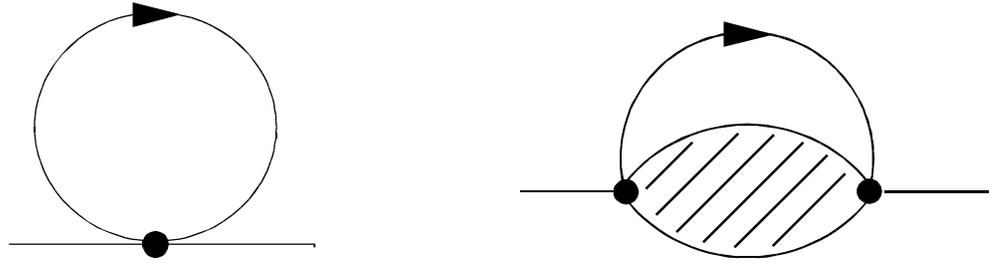
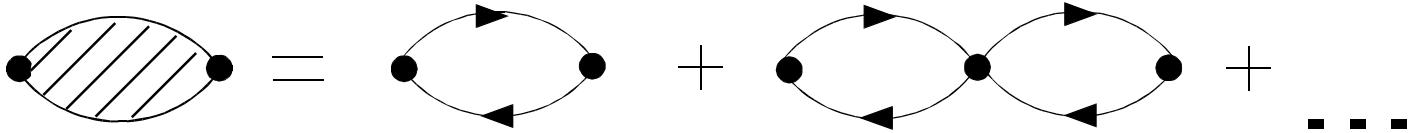
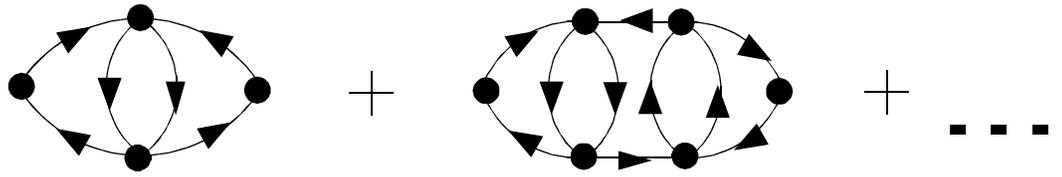
\end{document}